\documentclass[11pt,a4paper]{article}%%%%where rsproca is the template name
\usepackage[utf8]{inputenc}
\usepackage[margin=1in]{geometry}
\usepackage{graphicx}
\usepackage{amsmath}
\usepackage{amssymb}
\usepackage{hyperref}
\newcommand{\F}{\mathbb F}
\newcommand{\cF}{\mathcal F}
\newcommand{\cI}{\mathcal I}

\newcommand{\EE}{\mathbb E}

\newcommand{\PP}{\mathbb P}
\newcommand{\RR}{\mathbb R}

\newcommand{\C}{\mathcal C}
\newcommand{\T}{\mathcal T}
\newcommand{\cS}{\mathcal S}

\newtheorem{proposition}{Proposition}

\begin{document}

\title{Neutral phylogenetic models and their role in tree-based biodiversity measures}

\author{Mike Steel}
%\add{Biomathematics Research Centre, University of Canterbury, Christchurch, New~Zealand}

\maketitle
\begin{center}
{\em Biomathematics Research Centre, University of Canterbury, Christchurch, New Zealand}
\end{center}

\maketitle

\begin{abstract}
A wide variety of stochastic models of cladogenesis (based on speciation and extinction) lead to an identical distribution on phylogenetic tree shapes once the edge lengths are ignored. By contrast, the distribution of the tree's edge lengths is generally quite sensitive to the underlying model. In this paper, we review the impact of different model choices on tree shape and edge length distribution, and its impact for studying  the properties of phylogenetic diversity (PD) as a measure of biodiversity, and the loss of PD as species become extinct at the present. We also compare PD with a stochastic model of feature diversity, and investigate some mathematical links and inequalities between these two measures plus their predictions concerning the loss of biodiversity under extinction at the present. 
\end{abstract}


%%%%%%%%%% Insert the texts which can accomdate on firstpage in the tag "fmtext" %%%%%

\section{Introduction}

Stochastic tree models play a key role in phylogenetics at several levels. Firstly, they describe the process of speciation and extinction represented by phylogenetic trees, a topic that dates back to Yule's 1925 paper \cite{yul25}.  Further stochastic processes operate {\em within} the branches of  phylogenetic trees, for example to model the evolution of genetic sequences within species (under Markovian site substitution processes), or to describe the process of  incomplete lineage sorting of gene trees within a species tree (under the multispecies coalescent) \cite{kub23}. The investigation of these stochastic models has provided biologists with greatly enhanced techniques for inferring evolutionary trees from genomic data. Stochastic processes have also been applied to model lateral gene transfer (e.g. in prokaryotes) where genes move {\em between} the edges of a phylogenetic tree, with the goal of inferring a central tree-like signal from a large number of conflicting gene trees \cite{roc13}.  

%Here, we explore yet another process operating within  a tree: the gain and loss of features. 

In this paper, we focus on some generic properties of simple stochastic models that generate trees, and that operate within them. We first review some properties of neutral evolutionary models which induce a distribution on tree shapes that depends solely on the number of leaves. We describe some aspects of tree shape that depend on how speciation is viewed, and compare the predictions of the neutral model concerning tree shape with the corresponding predictions of the uniform model on phylogenetic trees.  We also describe some recent findings regarding the lengths of the edges (shortest, longest, average) in  birth--death trees. 

Next we focus on phylogenetic measures of biodiversity that are relevant to conservation biology theory (namely, phylogenetic diversity (PD) and feature diversity (FD)). The latter involves a birth--death type process operating within a (fixed or birth-death) tree.  We explore the relationship between these two measures on fixed trees and on classical birth--death trees, with particular attention to predicting the relative loss of diversity due to rapid extinction at the present. This includes some recent results, and possible questions for further investigation.

\subsection*{Types of trees}
\label{assec}

In this paper we deal with both fixed phylogenetic trees and random birth-death trees. A {\em phylogenetic tree} $T$ here refers to a rooted directed tree (having its edges directed away from the root) with  a set of labelled leaves (vertices of out-degree $0$) and with the remaining {\em interior} vertices unlabelled and of out-degree at least 2. If $X$ is the set of leaves of $T$ we say that $T$ is a {\em phylogenetic $X$--tree}.  A {\em star tree} is a phylogenetic tree with a single interior vertex. Mostly we deal with binary trees in which all interior vertices have out-degree exactly 2. Although we can view a phylogenetic tree as a (discrete) graph without edge lengths, we  will often also assign a (positive real-valued) length $\ell_e$ to each edge $e$ of the tree (and if these edge lengths are proportional to time they are said to be {\em ultrametric}). 
If we remove the labels on the leaves of a binary phylogenetic tree (and remove any edge lengths) we obtain a {\em tree shape} denoted $\tau$.   For example, when $|X|=4$, there are precisely two tree shapes: the balanced (`fork') shape and the unbalanced (`caterpillar') shape. 

\bigskip

\noindent\fbox{%
    \parbox{\textwidth}{%
        {\bf Notation:}
       Given a rooted phylogenetic $X$--tree $T$, an edge $e$ of $T$, and a leaf $x$ of $T$ we let:
\begin{itemize}
\item $n=|X|$; the number of leaves of $T$;
    \item $C_T(e)$ denote the set of leaves of $T$ that are descendants of edge $e$;
    \item $n_e = |C_T(e)|$; and
    \item $P_T(x)$ denote the set of edges on the path from the root of $T$ to leaf $x$.
\end{itemize}
    }%
}

\bigskip

\begin{figure}[htbp]
\begin{center}
\includegraphics[width=14cm]{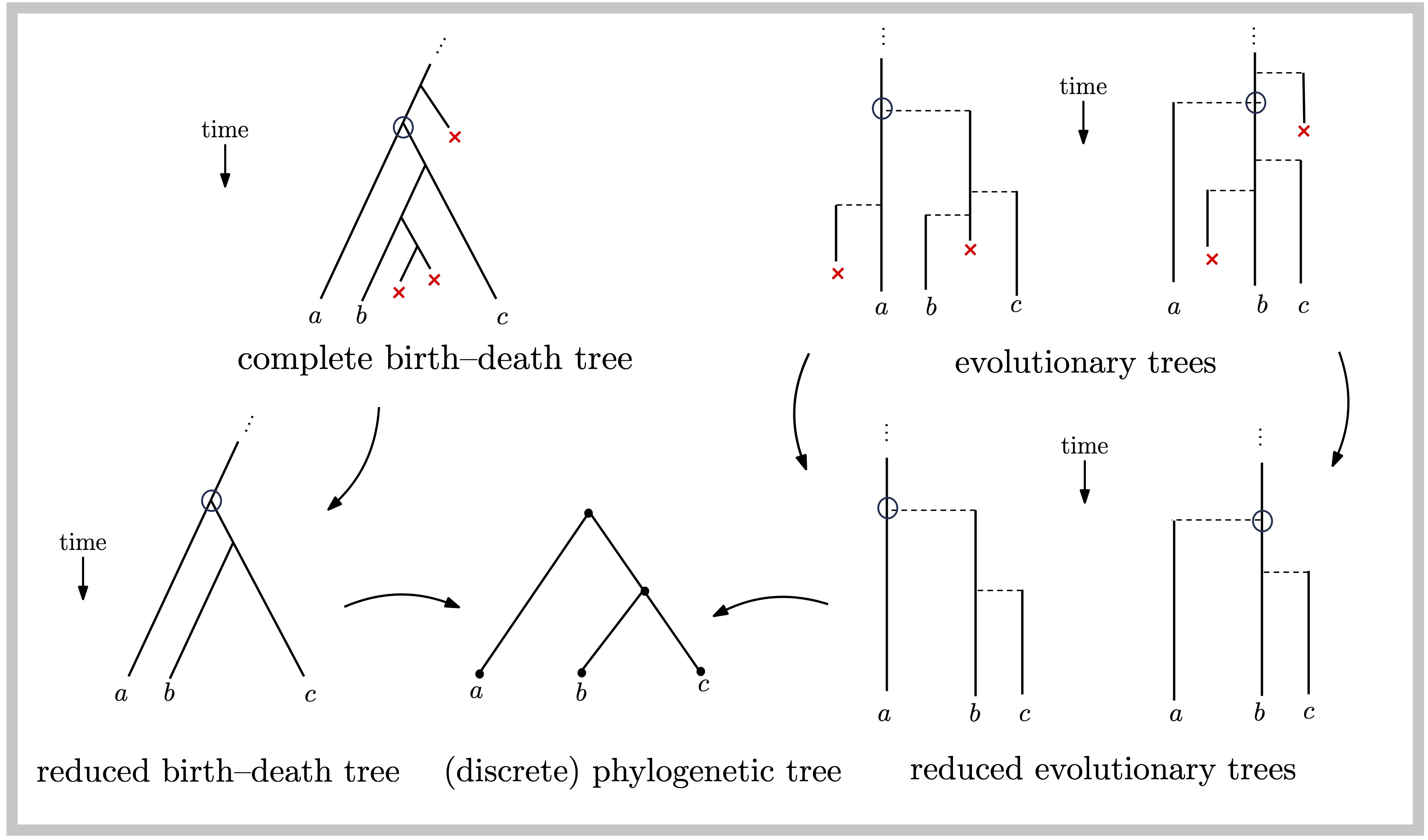}
\end{center}
    \caption{Classes of trees (see text for details).
} 
      \label{figfirst}
\end{figure}

A {\em birth--death tree} $\T_t$ is a rooted tree generated by a stochastic process of binary splitting and death of lineages, starting from a single lineage at time $0$ and acting for time $t$.  The tree $\T_t$ is {\em complete} if it includes lineages that become extinct by time $t$, whereas a {\em reduced} birth-death tree is obtained from the complete tree by removing these extinct lineages (this reduced tree is also referred to in the literature as the {\em reconstructed tree}). 
Notice that a reduced birth-death tree with extant species set $X$ induces a discrete phylogenetic $X$--tree (by ignoring the edge lengths, and the `stem' edge at the top of the tree), as indicated in Fig.~\ref{figfirst}(lower left). 
For either class of trees, a {\em pendant} edge in a tree is an edge incident with a leaf; all other edges are {\em interior}.

By convention, a vertex of out-degree 2 in a phylogenetic tree represents an ancestral speciation event.  This would typically involves a lineage splitting off from an existing lineage and forming a new species. However, there is usually no distinction made in a birth-death (or phylogenetic) tree as to which of the two edges corresponds to the formation of a `new' species, rather than a continuation of the existing one. This distinction has been noted by others (e.g. \cite{ald96, lam13a}, and invoking the asymmetry between parent and offspring lineages (as Yule did in his original model of genera evolution) can change how one measures various quantities.
%such as the number of times new species were generated in the path that leads to a given present-day species.  
To address this, we use the term  {\em evolutionary tree} for a tree that makes explicit which of the two edges directed out of a degree-2 vertex corresponds to the newly forming species (the horizontal edges in  Fig.~\ref{figfirst} upper (right)).

Evolutionary trees also  show lineages that lead to extinct taxa, and so we define the {\em reduced  evolutionary  tree}  to be the  tree obtained from an evolutionary  tree by (i) ignoring extinct lineages, and (ii) not treating a lineage as a new species if it branches off a lineage that has no surviving taxa at the present. This is illustrated  in Fig.~\ref{figfirst} (the two trees on the lower right). Once again, ignoring edge lengths in a reduced evolutionary tree leads to a discrete phylogenetic tree, as illustrated in Fig.~\ref{figfirst} (the circled vertices in the figure correspond to the root of the discrete phylogenetic tree).

%\newpage

\newpage
\section{Yule--Harding trees}
The classical birth--death process has constant birth and death rates ($\lambda$ and $\mu$, respectively) and has a long history \cite{yul25, fel39}. More general models were subsequently investigated  (e.g, \cite{ken48}) and which allow for these rates to depend on time, or the number of species present at a given time. In all such models,  the tree either becomes extinct or the number of leaves tends to infinity with probability 1 as $t \rightarrow \infty$ (e.g. by a more general result due to Jagers \cite{jag92}).  Moreover,  the shape of the resulting phylogenetic tree -- even allowing these rates to vary as described -- leads to the same distribution on phylogenetic tree shapes that  depends solely on $n$ (the number of leaves of the tree shape), the {\em Yule-Harding} (YH) distribution.

A very general model that leads to the YH distribution on tree shapes was described by Aldous (\cite{ald96}, p.8).  Briefly stated,  at any given time,  the probability that the next event (speciation or extinction) is a speciation rather than an extinction is allowed to depend arbitrary on the past. The only constraint required is that if the next event is a speciation (resp. extinction) then each species at that time is the equally likely to speciate (resp. become extinct).  We will refer to any such process of this type as a {\em neutral model}.
A further analysis of  models that  lead to the YH distribution on tree shapes has been described by Lambert and Stadler \cite{lam13a}.

\begin{figure}[!h]
    \centering
    \includegraphics[width=3in]{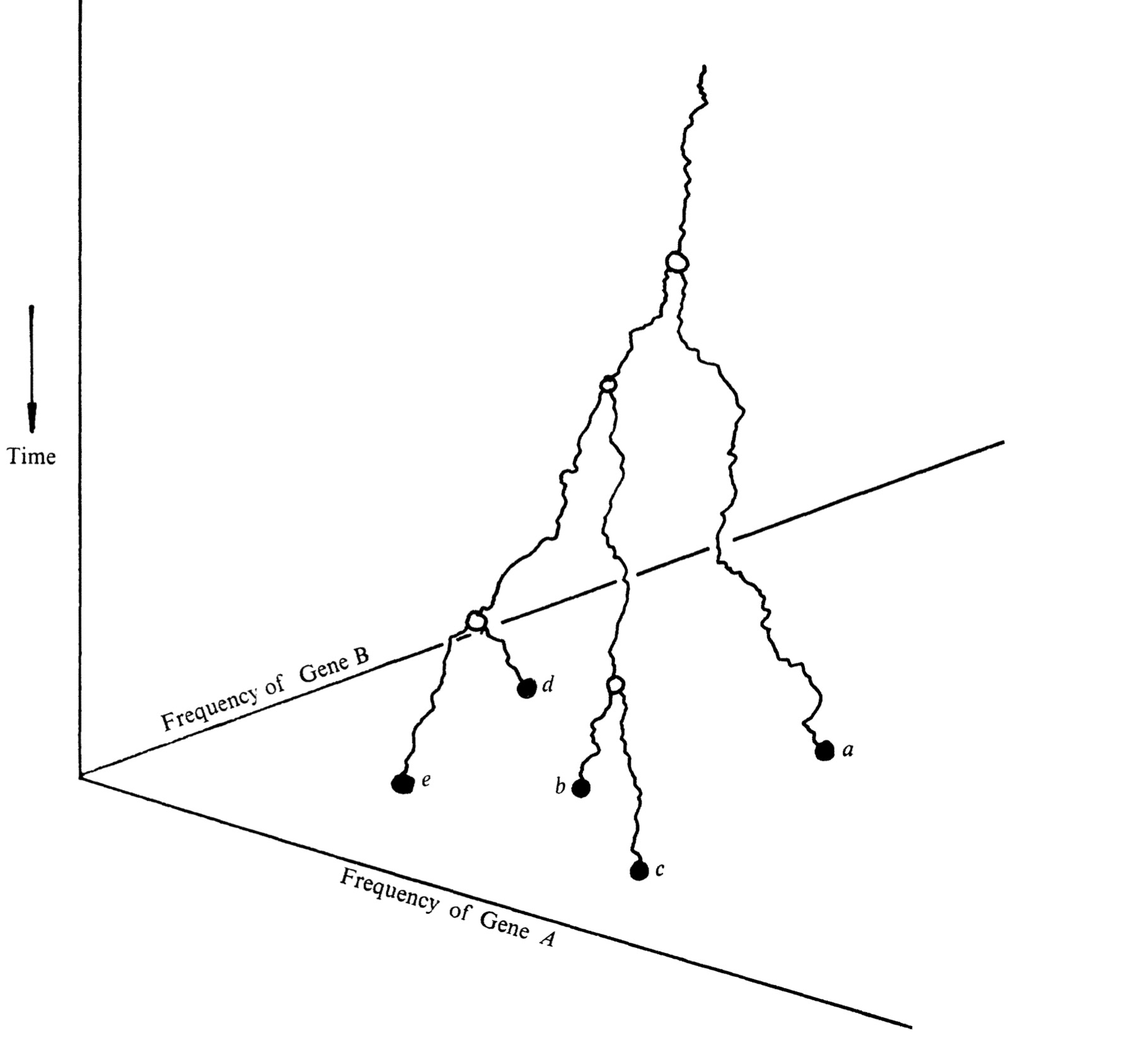}
    \caption{The binary phylogenetic tree from Harding's original 1971 paper \cite{har71}. After ignoring the edge lengths and leaf labels, there are three tree shapes on five leaves, and the tree shape shown has probability $1/6$. The tree itself (with leaf labels) has probability $1/90$.}
    \label{fig2}
\end{figure}

The following simple discrete stochastic process generates YH tree shapes: start with the tree on two leaves, and at each step, when the tree has $k \geq 2$ leaves,  select one of these leaves uniformly at random and attach a new leaf to the midpoint of its incident edge to produce a binary tree with $k+1$ leaves. If we continue until the tree has $n$ leaves, then the resulting tree shape has the YH distribution.  

For any tree shape $\tau$ with $n$ leaves there are precisely $n!/2^{s(\tau)}$ distinct binary phylogenetic trees with this shape, where $s(\tau)$ is the number of symmetry vertices in $\tau$ (i.e where the two descendent subtrees of the vertex have isomorphic shape). 
The probability $p(\tau)$ that the YH tree has a particular shape $\tau$ was described recursively by Harding \cite{har71}. However, it also  has an explicit formulation, namely: 
$$p(\tau) = \frac{2^{n-1-s(\tau)}}{\prod_{v \in IV_\tau}n_{\tau}(v)},$$
where  $s(\tau)$ is the number of symmetry vertices in $\tau$ (i.e., where the two descendent subtrees of the vertex have isomorphic shape), $n_{\tau}(v)$ is the number of leaves that are descendants of vertex $v$ minus 1, and $IV_\tau$ is the set of interior vertices of $\tau$. For example, the tree shape illustrated in Fig.~\ref{fig2} has probability $2^{5-1-3}/ (4 \times 3 \times 1 \times 1)= 1/6$.

For any tree shape $\tau$ with $n$ leaves there are precisely $n!/2^{s(\tau)}$ distinct binary phylogenetic trees with this shape. Thus, the  YH distribution on tree shapes induces a distribution on binary phylogenetic trees by multiplying  $p(\tau)$ by $2^{s(\tau)}/n!$. For example, the phylogenetic tree shown in Fig.~\ref{fig2}, has $n=5$ and $s(\tau)=3$, and so has probability $1/90$.

It turns out that the YH distribution also describes the distribution of tree shapes that arise under a somewhat different process, namely, the Kingman coalescent (again assuming that the edge lengths are ignored).

\bigskip

\noindent\fbox{%
    \parbox{\textwidth}{%
        {\bf Historical comment:}
        E.F. Harding investigated the enumeration and probabilities of tree shapes and phylogenetic trees more than 50 years ago in his seminal paper \cite{har71}. The enumeration of  tree shapes has a longer history, dating back to 1940, when Etherington \cite{eth39}  investigated the functional equation  $f(x)=x+\frac{1}{2}(f^2(x)+f(x^2))$ that describes the  generating function of these numbers. Harding also put forward  a conjecture regarding the shape of the most probable phylogenetic trees on $n$ leaves under this distribution. This conjecture was established a few years later \cite{ham74}. However, formulating a conjecture regarding the most probable tree shape on $n$ leaves (rather than the shape of a most probable phylogenetic tree, which has labelled leaves) eluded both Harding and the authors of \cite{ham74}, as the most probable shape exhibits an irregular behaviour as $n$ grows.
    }%
}

%In his 1971 paper, Harding put forward a conjecture regarding the shape of the most probable phylogenetic trees on $n$ leaves under this distribution. This conjecture was established a few years later \cite{ham74}. However, formulating a conjecture regarding the most probable tree shape on $n$ leaves (rather than the shape of a most probable phylogenetic tree) eluded both Harding and the authors of \cite{ham74}, as the most probable shape exhibits an irregular behaviour as $n$ grows.

\subsection{Depth of a random leaf in reduced trees}
\label{depsec}
The {\em depth} of an extant leaf $x$ in a reduced birth-death tree (or in a phylogenetic tree) is the number of edges between the root of the tree and leaf $x$. 
Similarly, the {\em depth} of an extant leaf $x$ in reduced evolutionary tree is the number of horizontal edges in the path from the root to $x$.  Let $D_n$ (resp. $D'_n$) denote the depth of a randomly chosen (extant) leaf in a reduced birth-death tree (resp. reduced evolutionary tree) with $n$ leaves.

Notice that for the reduced birth-death tree (or the phylogenetic tree) in Fig.~\ref{figfirst} the depth $D_3$ of a randomly chosen leaf  equals 2 with probability $2/3$ and equals $1$ with probability $1/3$. By contrast, in the reduced evolutionary trees in this figure there are two scenarios to consider. In the first (the left-hand reduced evolutionary tree in Fig.~\ref{figfirst}), the depth of a randomly chosen leaf is uniform on 0,1,2; in the other (the right-hand reduced evolutionary tree in Fig.~\ref{figfirst}) the depth is 0 with probability $1/3$ and 1 with probability $2/3$ (the right-hand reduced evolutionary tree in Fig.~\ref{figfirst}). Since the two scenarios are equally likely it follows that $D'_3 = 0,1, 2$ with probabilities $\frac{1}{3}, \frac{1}{2}$ and $\frac{1}{6}$, respectively. 

For neutral models (in which the reduced birth-death tree shapes follow the YH model), the distribution for $D_n$ has the following exact expression:\begin{equation}
\label{PPD1}
\PP(D_n = k) = 2^k \cdot \frac{c(n-1, k)}{n!};  \mbox{ for } k=1, \ldots, n-1,
\end{equation}
where $c(n-1, k)$ is the number of permutations on $n-1$ elements with $k$ cycles (the (unsigned) Stirling numbers of the first kind) \cite{mah92, lyn65}, and $\PP(D_n = k)=0$ for $k \geq n$ or $k=0$.

By comparison, for $D'_n$ under a neutral model  we have the following result (the short proof is provided in the Appendix):

\begin{proposition}
\label{pro1}
\mbox{For } k=0, \ldots, n-1,
\begin{equation}
    \label{PPD2}
\PP(D'_n = k) = \frac{c(n, k+1)}{n!},  
\end{equation}  
and $\PP(D'_n=k)=0$ for $k\geq n$.
Moreover, 
$\EE[D'_n] = \frac{1}{2}\EE[D_n]= \sum_{i=2}^n \frac{1}{i},$
and the total variational distance between $D_n'$ and a  Poisson distribution with mean $\ln(n)+\gamma-1$ converges to 0, where $\gamma$ is the Euler-Mascheroni constant.
\end{proposition}

\subsection{Yule--Harding trees vs. uniform phylogenetic trees}

We now list some of key features of the shape of Yule--Harding (YH) trees, and how they differ from the shape of binary phylogenetic trees sampled uniformly at random, which is often  called the {\em Proportional-to-Distinguishable-Arrangements} (PDA) model\footnote{ 
The YH model and PDA models are special cases of the $\beta$-splitting model of Aldous\cite{ald96}, with $\beta =0$ and  $\beta=-3/2,$ respectively.}.

Firstly, if $T_n$ is a YH tree, and we select uniformly at random one of the two subtrees incident with the degree-2 root of $T_n$ then the number $N$ of leaves  in this subtree tree is uniformly distributed between $1$ and $n-1$ (this  follows by a simple P{\'o}lya-Urn model argument). By contrast, the PDA model leads to a quite different distribution for $N$, since  the probability that one of the two subtrees incident with the root consists of just {\em one} leaf equals $n/(2n-3)$ which converges to $1/2$, as $n \rightarrow \infty$ (whereas the corresponding probability converges to $0$ under the YH model).

Another aspect of the shape of a tree is how `balanced' it is. There are many ways to measure tree balance (a recent book \cite{fis23a} provides an excellent survey).  However, tree balance, roughly speaking, usually measures the extent to which (on average across the tree) the two subtrees below the interior vertices differ in size.  
Empirical phylogenetic trees tend to be less `balanced' than the YH model predicts, but more balanced than expected in the  PDA model \cite{ald01}.

A further difference between the YH and PDA models concerns the depth of a randomly chosen leaf (recall that depth refers to the number of edges between the leaf and the root of the tree).   Suppose we select a leaf  uniformly at random from a YH tree $T_n$;  then the depth $D_n$ of this leaf  has an expected value of order $\ln(n)$ as $n \rightarrow \infty$ by eqn.~(\ref{PPD1}). By contrast, for the  PDA model, the corresponding expected value has order $\sqrt{n}$ \cite{ald96}.  

Next, given a YH  tree with $n$ leaves, if we select uniformly at random a pair of leaves, then their most recent common ancestor (MRCA) vertex is likely to be `close' to the root vertex of the tree. More precisely, the number of edges between the MRCA and the root converges in distribution (as $n$ grows) to a geometric random variable $G$ with $\PP(G=g) = \frac{1}{3}\cdot (\frac{2}{3})^g$ for $g = 0,1, \ldots$  \cite{ste01}. Thus, the probability that the MRCA vertex of the chosen pair of leaves is the root of the YH tree converges to $\frac{1}{3}$ as $n$ grows
\footnote{More generally, Sanderson \cite{san96}  showed that for a randomly selected subset of leaves $S$ of size  $k$, the probability that the MRCA of $S$ is the root of a YH tree converges to $1-\frac{2}{k+1}$ as $n \rightarrow \infty$.}.
For the PDA model however, the probability that  the   MRCA of the chosen pair of leaves is the root of the tree tends to zero as $n \rightarrow \infty$.

We end this section by describing a recent result that exhibits both a difference and a similarity between the YH and PDA model.
Let $\tau_k$ be the particular binary phylogenetic shape where all leaves have depth $k$ (thus $\tau_k$ has $2^k$ leaves).  

{\em Claim 1:} For each fixed $k$, the probability that the subtree of a YH tree $T_n$ consisting of all vertices of depth at most $k$ from the root has shape $\tau_k$ tends to $1$ as $n\rightarrow \infty.$

This claim follows by a slight extension of the argument used to establish Proposition 5 in \cite{bie24}.
Moreover, since every rooted tree shape is a subtree of $\tau_k$ for some (sufficiently large) $k$ it follows that the probability that a YH tree $T_n$ contains a subtree of shape $\tau$ tends to $1$ as $n \rightarrow \infty$. 
This last result also holds for the PDA model, but the proof is more involved (for details see \cite{bie24}) since Claim 1 (above) is easily shown to fail for the PDA model.

\subsection{Edge lengths of Yule trees}

Next, consider the {\em Yule process}, a continuous-time Markov process on state space $1,2,3 \ldots,$ with a linear birth rate $\lambda$ and a death rate of $0$. Such a process corresponds to the growing number of lineages in a random rooted binary tree in which each lineage of the tree splits independently according to an exponential distribution with rate $\lambda$. This tree is commonly referred to as {\em Yule tree} and we adopt this terminology here, despite it being somewhat contrived (as noted by Amaury Lambert in this special issue).

Suppose we now sample a Yule tree  just before the number of leaves of the tree changes from $n$ to $n+1$. Ignoring the initial stem edge, select any other edge of this tree uniformly at random from either (i) the set of pendant edges or (ii) the set of interior edges. In either case, the length of the sampled edge is exponentially distributed with expected value $\frac{1}{2\lambda}$. Thus the notion that the pendant edges might tend to be shorter, on average, than the interior edges (because the pendant edges are in some sense `still growing') is misleading. For further details, and extensions to birth-death trees see \cite{moo12}.
By contrast, if we select one of the two edges descended from the root of this Yule tree, then the expected length of this edge is different; it is precisely $\frac{1}{\lambda}\left(1-\frac{1}{n}\right)$ \cite{sta12}.

We can also consider the sum of the lengths of all the edges of a Yule tree (again ignoring the stem edge). Let $S_n$ denote this sum at the last moment when the tree has $n\geq 2$ leaves.  Since there are $2n-2$ edges in a binary tree with $n$ leaves, and since a randomly selected edge has expected length $1/2\lambda$, it follows that $S_n$ has expected value $(n-1)/\lambda$. However, more can be said:  $S_n$ has a Gamma distribution with variance $(n-1)/\lambda^2$,  and  is asymptotically normally distributed as $n \rightarrow \infty$ \cite{sta12}.

Moving beyond Yule trees, the distribution of the interior and pendant edge lengths in birth-death trees were investigated in \cite{moo12}, with more recent results reported in \cite{die24}.

\subsection{The longest pendant and shortest interior edges in birth-death trees}
We now consider the length of the {\em longest} pendant edge or the {\em shortest} interior edges. 
The former are of interest to biologists, since they end at a species that is maximally distant from the other taxa. The latter are relevant to tree reconstruction, as they represent near-polytomies that are hard to resolve using genomic data.
Assume $\lambda>\mu$. Let $L_t$ denote the length of the longest pendant edge in a reduced birth--death tree at time $t$, and let $N_t$ denote the number of leaves present at time $t$.  The following result is from \cite{boc23}.

\begin{proposition}
\label{pro3}
Conditional on $N_t>0$, the ratio $L_t /t$ converges in probability to $\frac{1}{2}$ and $\EE\left[ \frac{L_t}{t} | N_t>0\right]\to \frac{1}{2}$ as $\lambda t \rightarrow \infty$.
\end{proposition}
A noteworthy feature of Proposition~{\ref{pro3}} is that $L_t /t$ converges to a constant that involves neither $\lambda$ nor $\mu$. By contrast, the   longest pendant edge in the complete birth--death tree (i.e., including past extinctions)  divided by $t$ converges in probability to $\frac{1}{2}(1-\mu/\lambda)$ as $t$ grows \cite{boc23}.  A study across 114 mammal families  (with increasing numbers of leaves) provided a reasonable fit to the  value $t/2$  predicted by Proposition~\ref{pro3} \cite{boc23}, as did subsequent comparisons against other groups of taxa (e.g. birds) \cite{kom23}.
%For Yule trees ($\mu=0$), it is also possible to write $\EE[L_t/t]=\frac{1}{2} + \alpha(\lambda t)$ where $\alpha(x)$ is an explicit function (involving the arc-tan function), which converges to 0 exponentially fast as $x$ grows \cite{boc23}. 
A heuristic argument  for Proposition~\ref{pro3} was also described recently in \cite{dis23}, by considering ranked YH trees.  Results for the lengths of the $k$-th longest pendant edge are also possible.

It is also of interest to consider the length $S_t$ of the {\em shortest} interior edge, since resolving short edges from aligned DNA sequence data requires sequence lengths that (provably) grow with the inverse square of this length. Let $S_t$ denote the length of the shortest interior edge in a Yule tree $\T_t$ grown for time $t$. Then from \cite{boc23} (Proposition 4), we have: \begin{equation}
\label{shorty}
\PP\left(S_t \geq x e^{-\lambda t}/\lambda \right) \rightarrow   \frac{1}{1+ 2x}, \mbox{ as }  \lambda t \rightarrow \infty.
\end{equation}
Less precisely, Eqn.~(\ref{shorty}) with $x=\frac{1}{2}$ implies that, as $t$ grows, there is a $\sim$50\% probability that the shortest interior edge will be smaller than the average interior edge length divided by the expected number of leaves.

\section{Phylogenetic diversity}
\label{PDsec}
In the previous section, we considered the distribution of the sum of all the edge lengths in a Yule tree. This  sum can be viewed as some measure of the total amount of evolution that has taken place in the tree. More generally, for any phylogenetic tree $T$ on leaf set $X$  and any subset $Y$ of $X$, consider the sum of the lengths of the edges in the minimal subtree of $T$ that connects the leaves in $Y$ with the root of $T$, denoted here as $PD_{(T, \ell)}(Y)$. Formally, 
for an edge $e$ of $T$, we let $\ell_e$ denote any (real-valued and positive) `length' associated with $e$, and we let $\ell$ denote the function $e\mapsto \ell_e$ over all edges of $T$. Then
$$PD_{(T, \ell)}(Y) = \sum_{e: C_T(e) \cap Y \neq \emptyset} \ell_e.$$
This quantity, referred to as the {\em phylogenetic diversity} of $Y$, has been advocated as a measure of biodiversity, beginning with a seminal paper by Daniel P. Faith \cite{fai92}.  The function $Y \mapsto PD_{(T, \ell)}(Y)$  satisfies a combinatorial strong exchange property which allows a subset of any given size that maximises PD to be found via a  linear-time greedy algorithm.

Notice that the ratio $PD_{(T, \ell)}(Y)/PD_{(T, \ell)}(X)$  measures, roughly speaking, the proportion of the total edge lengths of $T$ that is spanned by $Y$, and is a quantity considered further in Section~\ref{pdbd}.  For birth-death trees we will assume that the edge lengths are proportional to time (and so ultrametric), however for fixed phylogenetic trees we allow arbitrary positive edge lengths.

\subsection{PD indices}
\label{pdindices}
Phylogenetic diversity provides a measure of diversity for subsets of species. However, conservation biologists are also  interested in quantifying how much each species contributes to the total PD of a phylogenetic $X$-tree $T$. In other words, they are interested in functions $I_{(T, \ell)}: X \rightarrow \RR^{\geq 0}$ for which $$\sum_{x \in X}I_{(T, \ell)}(x) = PD_{(T, \ell)}(X).$$ 
We also require that $I_{(T, \ell)}(x)$ is a linear function of the edge lengths (thus, we can write $I_{(T, \ell)}(x) = \sum_{e}c_e(x)\ell_e$ for values $c_e(x)$) and, moreover, that $c_e(x) = 0$ is $x$ is not a descendant of $e$, with $c_e(x)$ is strictly positive otherwise. 
The class of such functions can thus be written as
\begin{equation}
    \label{eqindex}
I_{(T, \ell)}(x) = \sum_{e \in P_T(x)}c_e(x) \ell_e,
\end{equation}
 and they are referred to as {\em PD indices}.

The most popular PD index is {\em fair proportion (FP)} (also called `evolutionary distinctiveness'), introduced in \cite{red08}, and defined by putting $c_e(x) = 1/n_e$  in eqn. (\ref{eqindex}). In other words, $FP$ shares out the length of each edge equally amongst the leaves of $T$ that are descendants of that edge.  This `fair sharing' property is consistent with the mathematical  result that FP is precisely the Shapley value of a cooperative game that is naturally associated with any phylogenetic tree with edge lengths \cite{fuc15}.  Another PD index is  {\em equal splits} ($ES$) defined by putting $c_e(x) = 1/N_e(x)$  in eqn. (\ref{eqindex}) where $N_e(x)$ is the product of the out-degrees of the interior vertices of $T$ in the path from $e$ to $x$.  Other PD indices exist (e.g. by taking any convex combination of FP and ES). 
%The set of all PD indices for a given tree $T$ forms a convex space (i.e., if $I_{(T, \ell)}$ and $I'_{(T, \ell)}$ are diversity indices for $T$ then so is $\alpha I_{(T, \ell)} + (1-\alpha)I'_{(T, \ell)}$, for any $\alpha \in [0,1]$). The collection of diversity indices is larger than the convex space spanned by FP and ES, except for small values of $n$ \cite{man23}.
 
Although a PD index satisfies the equation $\sum_{x \in X}I_{(T, \ell)}(x) = PD_{(T, \ell)}(X)$, if we consider a subset $Y$ of $X$, the value $\sum_{x \in Y}I_{(T, \ell)}(x)$ does not generally equal $PD_{(T, \ell)}(Y)$, except in special cases (e.g. if  $T$ is a star tree). Nevertheless, a general inequality holds for any phylogenetic $X$-tree $T$, namely: 
$$\sum_{x \in Y}I_{(T, \ell)}(x) \leq  PD_{(T, \ell)}(Y), \mbox{ for all } Y \subseteq X.$$
This result is from \cite{bor24} (Lemma 1.1), where the authors explored how large the difference
$PD_{(T, \ell)}(Y) - \sum_{x \in Y}I_{(T, \ell)}(x)$ can be (over subsets of leaves $Y$ of $X$) when $I_{(T, \ell)}$ is the FP or ES index and $T$ is either fixed or free (and $\ell$ fixed or constrained).
A further recent result concerning PD indices is that the rankings of species by FP can change (and even reverse) if some of the  species were to disappear from the tree (for details, see \cite{fis23, man24}). 

Turning next to random trees, let  $FP_{\T_t}(x)$ be the fair proportion index of a leaf selected uniformly at random from a Yule tree $\T_t$ grown for time $t$.   In this case,
$$\EE[FP_{\T_t}(x)] = \frac{1}{\lambda} (1+e^{-\lambda t}),$$
(from \cite{boc23} equation (32)) and so $FP_{\T_t}(x)$ converges in expectation to $\frac{1}{\lambda}$ as $t$ grows. Since the expected length of the pendant edge incident with $x$ converges to $\frac{1}{2\lambda}$ as $t$ grows, one expects around half of the FP index for a random leaf to be contributed by its pendant edge, which is consistent with an early empirical observation from \cite{red08}.  

PD indices have also been recently defined and studied in a more general setting that includes unrooted phylogenetic trees, phylogenetic networks, and set systems \cite{wik20,nil24,mou24}.

\subsection{The loss of PD under extinction at the present}
Mass extinction events have played an important role in evolution \cite{rau93} and we are currently entering a new (anthropogenic) one. The simplest process  of extinction at the present is the {\em field-of-bullets} (FOB) model, in which species become extinct independently, and with the same probability $\epsilon$. A more nuanced process -- the {\em generalised field-of-bullets} model (gFOB) -- assigns each species its own extinction probability  $\epsilon_x$ while still assuming independence. In practice, species extinction probabilities can be estimated from their Red List status (from the International Union for Conservation of Nature) \cite{moo08, gum23}.  More realistic extinction processes would account for the fact that the extinction of some species may trigger the further extinction of other species (leading to `extinction cascades'); alternatively, species that share common features may be correlated in their extinction risk if these features turn out to be deleterious. 
We do not consider these more complex models here. 

For a given phylogenetic $X$-tree $T$ with given edge lengths $\ell$, the expected proportion of PD that is lost under a gFOB model is given by:   
\begin{equation} 
\frac{\sum_e \ell_e \prod_{x \in C_T(e)} \epsilon_x}{\sum_e \ell_e}
\label{losseq}
\end{equation}
To justify (\ref{losseq}) simply observe that $\ell_e$ fails to contribute to the post-extinction PD score  precisely if all species descended from $e$ become extinct.
Provided that the distribution of the edge lengths is not too extreme, and the $\epsilon_x$ values are bounded away from 0 and 1, the distribution of PD on $T$ that is lost becomes asymptotically normally distributed as the number of leaves grows (for details, see \cite{fal08}, Theorem 3.1). Next we consider the loss of PD for birth--death trees (rather than for fixed trees).

\subsection{Loss of PD on birth--death trees under extinction at the present}
\label{pdbd}

\begin{figure}[htb]
\centering
\includegraphics[scale=0.07]{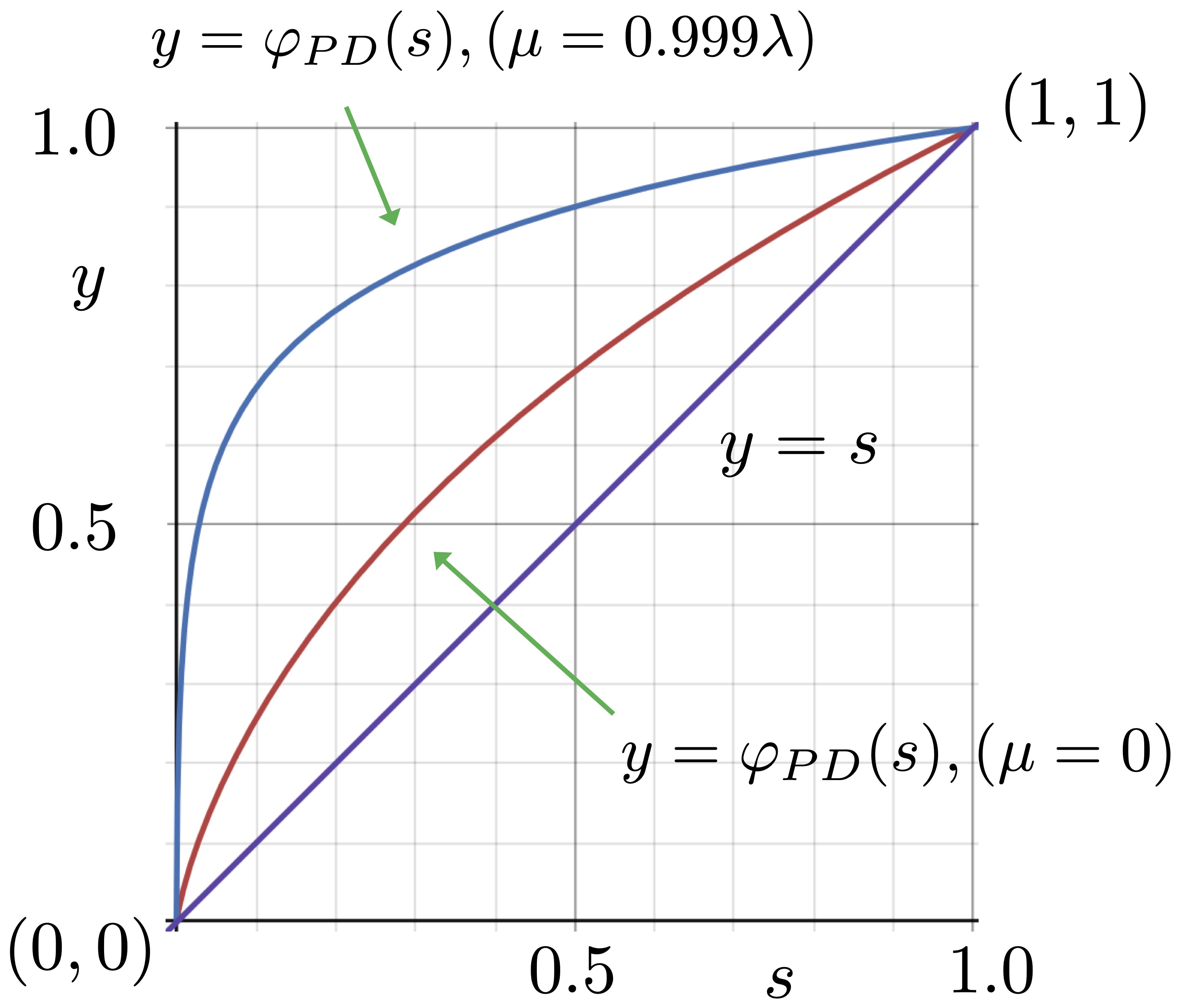}
\caption{The proportion of PD remaining (y-axis) after a field of bullets extinction of variable intensity (x-axis is the survival probability $s$) at the tips of a birth--death tree as $t \rightarrow \infty$. Except when $s =0,1$ this proportion always exceeds the proportion of species remaining (bottom line) and increases from the curve corresponding of a Yule tree (middle curve) to the unit step function as the birth--death process approaches criticality ($\mu/\lambda \rightarrow 1$).}
\label{figgraph}
\end{figure}

Let $\T_t$ be a birth--death tree  with $\lambda\geq \mu$, and apply the simple FOB model, with survival probability $s$, to the extant leaves of $\T_t$.  What proportion of the tree remains? More precisely, let $X_t$ denote the extant leaves of $\T_t$ and let $Y_t$ denote the subset of $X_t$ of the species that survive the FOB extinction event. What then can be said about the proportion of PD that survives, or, in other words, the ratio $PD_{\T_t}(Y_t)/PD_{\T_t}(X_t)$?  From \cite{lam13}  the following convergence in probability holds as $t$ becomes large:
$$\frac{PD_{\T_t}(Y_t)}{PD_{\T_t}(X_t)} \xrightarrow[t \rightarrow \infty]{P} \varphi_{PD}(s)$$
where 
 $$\varphi_{PD}(s) :=\lim_{t \rightarrow \infty} \frac{\EE[PD_{\T_t}(Y_t)]}{\EE[PD_{\T_t}(X_t)]}.$$
 When $\mu=0$ (i.e., a Yule tree),  $\varphi_{PD}(s)$ has a particularly simple form, namely:

\begin{equation}
\label{pieqyule}
    \varphi_{PD}(s) = \frac{s\ln(1/s)}{1-s}.
\end{equation}
Otherwise, if $\mu >0$, let $\rho = \mu/\lambda$, in which case:
\begin{equation}
\label{pieq}
\varphi_{PD}(s): = \begin{cases} 
\frac{\rho s}{\rho +s-1} \cdot \frac{\ln(s/(1-\rho))}{\ln(1/(1-\rho))}, & \mbox{if } \rho= \mu/\lambda \neq 0, 1-s;\\
 (1-s)/\ln(1/s), & \mbox{ if $\rho =1-s$}.
 \end{cases}
\end{equation}
In all cases, $\varphi_{PD}(s)$ is a monotonically increasing concave function that satisfies $\varphi_{PD}(s)\geq s$ with equality if and only if $s\in \{0,1\}$. This is shown in Fig.~\ref{figgraph}, which also illustrates an interesting  feature of $\varphi_{PD}(s)$: as $\mu$ approaches $\lambda$ from below, the function $y=\varphi_{PD}(s)$ approaches the unit step function (which is zero at $s=0$ and is $1$ for $s \in (0,1]$). Roughly speaking, this is because the resulting critical trees, conditional on non-extinction, have very short pendant and near-pendant edges. 

Moreover, for any value of $\mu$ and $\lambda$, as the survival probability ($s$) approaches 1 the function $y=\varphi_{PD}(s)$ has a slope equal to the expected proportion of the PD of the tree that is contributed by its pendant edges. The intuition behind this is that when $s$ is close to 1, the only edges that disappear are isolated pendant edges (interior edges only start to disappear when two adjacent leaves disappear). Thus, for Yule trees, the slope of $y=\varphi_{PD}(s)$ at $s=1$ is $\frac{1}{2}$, whereas for birth--death trees with $\mu>0$, the slope at $s=1$ is smaller.
%(for further details, see \cite{lam13}).

\bigskip

\bigskip

\noindent\fbox{%
    \parbox{\textwidth}{%
        {\bf Heuristic derivation of $\varphi_{PD}$: }
  To provide some intuition for eqn. (\ref{pieqyule}), here  we give a brief (but non-rigorous) derivation of it.
   First recall that constant-rate birth-death trees induce a tree shape that follows the YH distribution, and for this distribution, the expected number of edges that have $k\leq n-1$ leaves as descendants is $2n/(k(k+1))$. Next, observe that the length of each such edge contributes to the PD score with probability $(1-(1-s)^k)$ (i.e. iff at least one of these $k$ leaves survives). Now for a Yule tree, the expected length of the (interior and pendant) edges is the same, and so $\varphi_{PD}(s)$ reduces to $\sim \sum_{k=1}^n \frac{1}{k(k+1)} (1-(1-s)^k)$ which converges to  $s\ln(1/s)/(1-s)$ as $n \rightarrow \infty$.
    }%
}

\subsection{Stochastic diversity indices for extinction at the present}
Let $X$ denote a set of extant species, and consider the (random) subset $\cS$ of species in $X$ that will still be extant (i.e. still present) at some future time $t_*>t$ under some extinction model. Let $(T, \ell)$ be an associated phylogenetic tree with leaf set $X$. 
Focusing on any particular species $x \in X$, let
 $\cS_x = \cS \setminus\{x\}$ and let 
$$\psi_x = \EE[PD_{(T, \ell)}(\cS_x \cup \{x\}) - PD_{(T, \ell)}(\cS_x)].$$
Thus $\psi_x$ measures the expected additional PD at time $t_*$ if species $x$ is present compared with if it is absent at time $t_*$ (noting that other species may also be extinct at time $t_*)$. If we now specialise the extinction model to the gFOB model, where each species $s$ in $X$ has probability $\epsilon_s$ of being extinct at time $t_*$, then:
\begin{equation}
\label{eq1}
\psi_x = \sum _{e \in P_T(x)} \ell_e \prod_{s \in C_T(e)\setminus\{x\}} \epsilon_s
\end{equation}
Analogously, let
$$\psi'_x = \EE[PD_{(T, \ell)}(\cS_x \cup \{x\}) - PD_{(T, \ell)}(\cS_x\cup \cI_x)],$$
where $\cI_x \in \{\emptyset, \{x\}\}$, and $\epsilon_x = \PP(\cI_x = \emptyset)$.
Thus, $\psi'_x$ is the expected additional  PD at time $t_*$ if species $x$ is present at time $t_*$ compared with if its state is uncertain at time $t_*$.
Straightforward algebra leads to the following relationship between $\psi_x$ and $\psi'_x$ under the gFOB model:
\begin{equation}
\label{edge}
\psi'_x = \psi_x \cdot \epsilon_x.    
\end{equation}
(see \cite{ste07} for details).
The  indices $\psi$ and $\psi'$ linked by equation~(\ref{edge}) are part of the recently-proposed EDGE2 protocol \cite{gum23} for assessing potential biodiversity loss in coming decades. 
Note that $\psi$ and $\psi'$ are not PD indices in the sense of Section~\ref{pdindices} since summing either of them over all elements of $X$ is not, in general, equal to $PD_{(T, \ell)}(X)$ (however this is the case for $\psi$ when $T$ is a star tree).

We end this section by noting a curious connection between the index $\psi$ and the fair proportion index from Section~\ref{pdindices}. Under the FOB model, suppose that the extinction probability $\epsilon$ is described by a uniform distribution on $[0,1]$. Then:
\begin{equation}
\label{link}
    FP_{(T, \ell)}(x) = \EE[\psi_x(\epsilon)], \mbox{for all } x \in X,
\end{equation}
where the expectation is with respect to this uniform prior on $\epsilon$.  The short proof of Equation~(\ref{link}) is in the Appendix. 

\section{Feature diversity (FD) - a generalisation of PD}
\label{FDsec}

The PD of a subset $Y$ of extant species is sometimes assumed to be a proxy for the diversity of features (traits, characteristics, etc.) that are present among the species in $Y$. 
Under certain settings, this can be shown to be formally correct, as we describe shortly, however, under more realistic models of feature evolution it does not exactly apply. One reason, which we model here, is that features are not only gained along lineages, but can also be lost (e.g. the evolution of flight in birds which has subsequently been lost in some species). Indeed, a number of authors have pointed out that PD can be a poor proxy for feature diversity (FD) (e.g. \cite{maz18,kel14}). 

Features can be either discrete or continuous. Here, for simplicity, we consider just the discrete setting. Suppose that each species $x$ in $X$ has an associated set $F_x$ of features (some or all of which may be shared by other species).   Let $\F = (F_x: x \in X)$ denote the assignment of features to each species in $X$, and suppose that  feature $f$ has an assigned strictly positive real-valued weight $w(f)$ (this may reflect the complexity or importance of this feature).  For any subset $Y$ of $X$, its {\em feature diversity}, denoted $FD_\F(Y)$, is  given by:
$$FD_\F(Y) =\sum_{f \in \bigcup_{y \in Y}F_y} w(f);$$
(i.e. the sum of the $w(f)$ values across all features present in at least one species in $Y$). Thus, when $w(f)=1$ for all features then $FD_\F(Y)$ is simply the number of features present among the species in $Y$. 
 Notice that, unlike PD, FD does not involve any phylogenetic tree.  Nevertheless, we can view PD as a special case of FD (but not conversely) as the following box explains.

\bigskip

\noindent\fbox{%
    \parbox{\textwidth}{%
        {\bf PD as FD:}
      Given $(T, \ell)$, PD can be modelled exactly by FD as follows: Assign to each species $x$ in $X$ a feature $f_e$ for each edge $e$ on the path in $T$ from the root to leaf $x$, and assign feature $f_e$ weight $w(f_e)=\ell_e$.  Then $PD_{(T, \ell)}(Y) = FD_\F(Y)$ for all $Y\subseteq X.$ This abstract identity can be helpful. For example,  the function $Y \mapsto FD_\F(Y)$ (for any $\F$) satisfies the  submodular inequality (which states that $\phi(Y_1 \cup Y_2)+\phi(Y_1 \cap Y_2) \leq \phi(Y_1)+\phi(Y_2)$) and so PD does too.  However, FD cannot generally be represented by PD.  A counterexample is the feature sets for species $x,y,z$ given by $F_x= \{f_1, f_2\}, F_y = \{f_1, f_3\}$ and  $F_z=\{f_2, f_3\}$ with $w(f_i)=1$). The non-nested nature of these sets implies that the resulting FD function cannot be realised as PD on any tree with edge lengths.
    }%
}

\bigskip

It turns out that FD can be represented by PD on some tree when the collection $\C_\F$ of  subsets $X_f =\{x \in X: f \in F_x\}$ is nested (i.e., the condition that $A, B \in \C_\F \Rightarrow A \cap B \in \{A, B, \emptyset\}$ holds), by results in \cite{wic21} (Proposition 1 and Theorem 1). Moreover, since a nested collection of subsets of $X$ corresponds to a subset of the clusters of a  phylogenetic tree, it is possible to describe the evolution of each character in $\F_X$ on that tree by a single-gain (of the feature on one edge $e$) and no subsequent loss (on any descendant edge of $e$).  A slightly stronger statement (from \cite{wic21}) is the following:
\begin{proposition}
\label{pro4}
    For feature assignment $\F_X$, there is a phylogenetic $X$-tree $T$ with non-negative edge lengths $\ell$ (not necessarily ultrametric) so that $FD(Y) = PD_{(T, \ell)}(Y)$ for all subsets $Y$ of $X$ if and only if the evolution of each feature in $\F_X$ can be described by an evolutionary scenario on $T$ involving single-gains and no-loss.
\end{proposition}
 In fact, the `only if' part holds even if we restrict the subsets $Y$ of $X$ to be of sizes of at most 3. The key phrase in Proposition~\ref{pro4} is ``can be described"; there is no requirement that this is how the feature actually evolved. For example, a feature that arises on an interior edge, and is subsequently lost on one of the edge's two descendant edges, can be described by a scenario that involves a single gain (on the other child edge)  and no loss.

\subsection{A stochastic model of FD on fixed trees}
\label{stocsec}
\begin{figure}[htbp]
\begin{center}
\includegraphics[width=15cm]{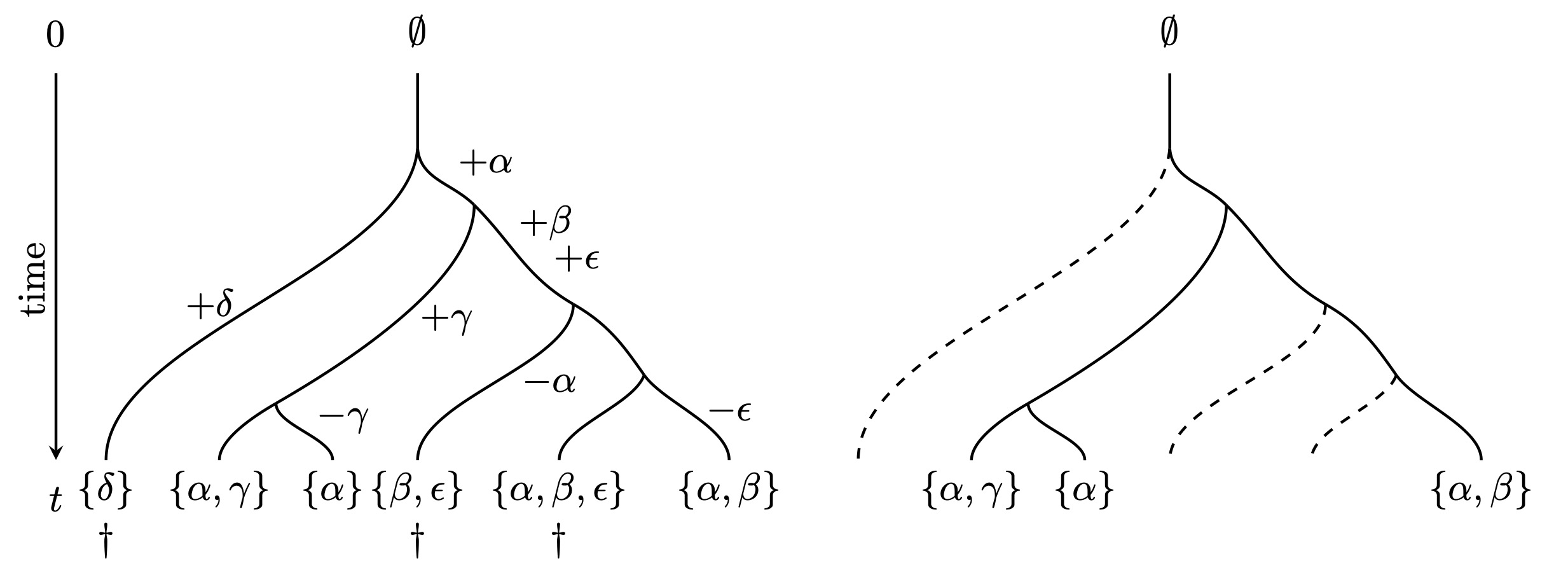}
\end{center}
    \caption{A simple example to illustrate the gain and loss of features on a tree and the FOB process (from \cite{ove23}). {\em Left:} New features arise (indicated by $+$) and disappear (indicated by $-$) along edges of the tree. Each new features arises only once in the tree, and for convenience, no features are present at time $0$.  Five features are present among the leaves of  the tree at time $t$:  $\{\alpha,\beta,\gamma,\delta,\epsilon\}$. A rapid extinction event at the present then results in the loss of three leaves of the tree (denoted by $\dagger$). {\em Right:} Three features ($\alpha,\beta,\gamma$) remain among the leaves of the resulting tree. Thus, the ratio of surviving features to total features is $3/5= 0.6$.
} 
      \label{figPDFD}
\end{figure}

 Proposition~\ref{pro4} is purely combinatorial,  so it is instructive to consider a stochastic model of feature evolution on a phylogenetic $X$-tree $T$ (with positive edge length function $\ell =[\ell_e]$), where each new feature arises once in the tree and can subsequently be lost. This is illustrated in Fig.~\ref{figPDFD}.

 In this model, 
 \begin{itemize}
     \item a new feature arises (on any edge) at rate $r>0$ per unit time,  and
     \item each feature (on any edge) is subject to an independent loss process with rate $\nu \geq 0$.
 \end{itemize}
This simple model can be  described by a classical (constant-birth, linear-death) stochastic process for which the following results hold (see e.g. \cite{fel50}, p. 461).
 Consider a directed path in the tree $T$ of length $L$ (the sum of the $\ell_e$ values on the path).  Conditional on there being $k$ features present at the start of this path,  let $N_L^{(k)}$ denote the number of features present at the end of this path. For $k=0$, $N_L^{(0)}$ has a Poisson distribution with mean  $\frac{r}{\nu}(1-e^{-\nu L})$, and this mean is 
 asymptotic to $rL$ as $L \rightarrow 0$ and to $r/\nu$ as $L \rightarrow \infty.$  More generally, if there are $k$ features present at the start of the path, then $N_L^{(k)}$ is the sum of two independent random variables:  one distributed as $N_L^{(0)}$,  the other has a Gamma distribution with parameters $(k, \nu L)$.  In particular, as $L \rightarrow \infty$, $N_L^{(k)}$ converges to a Poisson distribution with mean $r/\nu$.

To simplify what follows, we will assume that $w(f)=w>0$, though similar results  hold more generally  when $w(f)$ are i.i.d. random variables taking values in  $[0, B]$ for some constant $B$. 
Given a phylogenetic $X$-tree $T$, the assignment $\F_X$ of features present at the leaves of $T$ is now a random variable, and so is $FD(Y)$ for each subset $Y$ of $X$.  Let us then consider the expected FD value (i.e., $\EE[FD_{(T, \ell)}(Y)]$) which depends on $T, \ell=[\ell_e], r$ and $\nu$. 

\begin{figure}[htbp]
\begin{center}
\includegraphics[width=10cm]{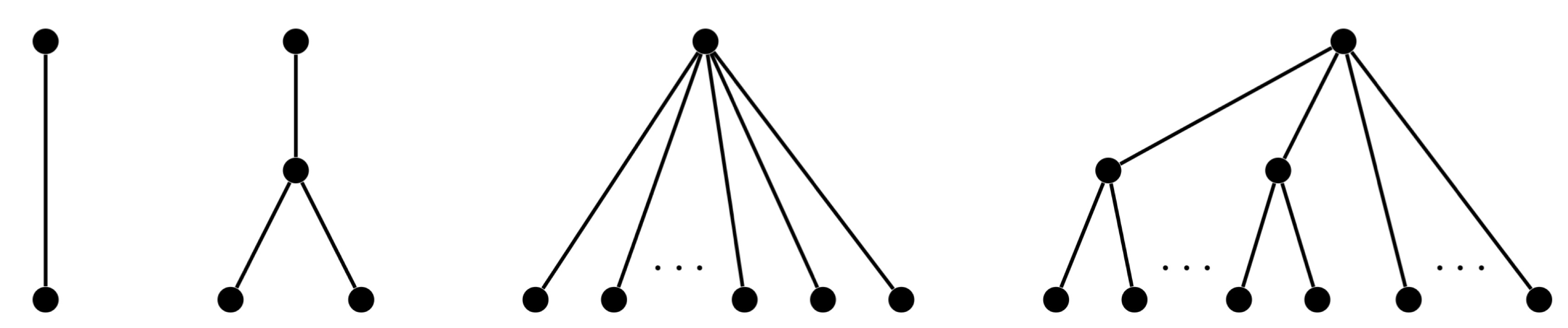}
\end{center}
    \caption{The  (only) four shapes of trees for which the expected feature diversity can be exactly described by phylogenetic diversity (from \cite{ros23}).
} 
      \label{fig_4trees}
\end{figure}

We can now ask whether the expected FD function $Y \mapsto \EE[FD_{(T, \ell)}(Y)]$ coincides exactly with the PD function on $T$, with some appropriately chosen non-negative edge lengths (say $\ell'$). 
It is easy to see that when the rate of loss $\nu$ equals zero, then the two functions always coincide (and with $\ell =\ell'$). Moreover, when $\nu>0$ and $T$ is one of the four tree shapes shown in figure~\ref{fig_4trees}, the expected FD function can be described exactly by PD for an appropriate choice of $\ell'$. Note that these four classes of trees all have depth at most 2; however, not all trees of depth 2 fall into one of these four classes.
Apart from these special cases, the following result was recently established in \cite{ros23} (Supplementary Material, Theorem 1.5). It can be regarded as a stochastic analogue of the combinatorial result in Proposition~\ref{pro4}.

\begin{proposition}
\label{pro5}
Suppose that $\nu>0$, and $T$ is a phylogenetic $X$-tree, with strictly positive edge lengths $\ell$. Then there exists edge lengths $\ell'$ for $T$ for which $\EE[FD_{(T, \ell)}(Y)] = PD_{(T, \ell')}(Y)$ for all subsets $Y$ of $X$ if and only if $T$  has the shape of one of the four trees shown in Fig.~\ref{fig_4trees}.
\end{proposition}

As was the case for Proposition~\ref{pro4}, the `only if' part of Proposition~\ref{pro5} also holds if we restrict the subsets $Y$ of $X$ to be just those of size at most 3. This result remains true even if  one allows the edge lengths $\ell$ and $\ell'$ to be non-ultrametric.

The function $Y \mapsto \EE[FD_{(T, \ell)}(Y)]$ (and a related deterministic measure of the evolution of feature diversity) forms the basis of the recently proposed `EvoHeritage' suite of biodiversity measures \cite{ros23}, in which the gain and loss of features provides a more nuanced measure of biodiversity than simpler PD-based measures.

\subsection{FD in birth--death trees with extinction at the present}

In the previous section we considered FD on a given tree under a stochastic model of feature gain and loss. In this section we allow the tree to be random (a birth-death tree) and subject to the FOB model operating on the leaves at the present. Thus we have  the following three-layer stochastic process:
\begin{itemize}
    \item A birth--death tree $\T_t$ with parameters $\lambda,\mu$ grown for time $t$ from time 0;
    \item A feature evolution model operating on the edges of $\T_t$ with gain and loss parameters $r, \nu$ (as in Section~\ref{stocsec});
    \item  A FOB model operating at time $t$ on the tips of $\T_t$,  with survival probability $s$.
\end{itemize}  

In this setting, we are interested in determining the proportion of features at the tips of the tree that survive the FOB model.

Next, we recall a standard result from birth--death theory. 
Consider a linear birth--death  process $X_t$, with birth rate $\lambda$, death rate $\theta$,  starting from a single individual at time $0$. At time $t$,  sample the individuals present independently with sampling probability $s>0$, and let  $X_t$ ($ t\geq0$) denote the number of these sampled individuals, and let $R_t^s(\lambda,\theta)= \PP(X_t>0).$
In this case:
\begin{equation}
\label{eqw}
R_t^s(\lambda,\theta)=\begin{cases}
\frac{s(\lambda-\theta)}{s\lambda+(\lambda(1-s)-\theta)e^{(\theta-\lambda)t}}, & \mbox{ if } \lambda \neq \theta;\\
\frac{s}{1+\lambda s t}, & \mbox{ if } \lambda = \theta.
\end{cases}
\end{equation}
In particular,  $R_t^s(\lambda,\theta)$ converges to 0 if $\lambda \leq \theta$ and  converges to a strictly positive value, namely $1-\theta/\lambda$, if $\lambda> \theta$ \cite{ken48, yan97}.

Let $\T_t^s$ denote the tree obtained from $\T_t$ by deleting all leaves at time $t$ that do not survive the FOB model. The  number of species at time $t$ in $\T_t^s$ that have a copy of a particular  feature  that arose at some fixed time $t_0 \in (0, t)$ in $\T_t$ is described exactly by the birth--death process $X_{t-t_0}$ with birth rate $\lambda$, death rate $\theta = \mu+\nu$ and extinction at the present (with survival probability $s$ for each individual).
It  follows that as $t$ becomes large, it becomes increasingly certain that none of the species at time $t$ in $\T_t$ will contain a given feature $f$ present at $t_0$ if $\lambda \leq \mu+\nu$, whereas if $\lambda> \mu+\nu$, there is a positive limiting probability that $f$ will be present in the extant leaves of the $\T_t$ (even after pruning under the FOB model).

Let  $\cF_t$ (respectively $\cF^s_t)$ be the random variable that counts the number of features present among the leaves of $\T_t$ before (respectively, after) a FOB extinction event at the present (with a survival probability per species of $s$). Thus,  $\EE[\cF_t] = \EE[\cF_t^1]$, and we have:
\begin{equation}
\label{FFT}
    \EE[\cF_t^s] = \int_0^t re^{(\lambda-\mu)\tau}R^s_{t-\tau}(\lambda, \mu+\nu) d\tau +F_0 R^s_t(\lambda, \mu+\nu),
\end{equation}
where $F_0$ is the number of features initially present at time $0$ (i.e., at the top of the stem edge).

An intuitive (but not rigorous) justification of eqn. (\ref{FFT}) is as follows. The expected number of tree lineages at time $\tau$  is $e^{(\lambda-\mu)\tau}$, so the expected rate at which new features arise at time $\tau$ is $r$ times this function. We then multiply this expected rate by the probability that a given feature at time $\tau$ survives to the present and also avoids the FOB extinction (i.e., we multiply by $R^s_{t-\tau}(\lambda, \mu+\nu)$), and integrate this modified rate from $0$ to $t$ to obtain the first term in eqn. (\ref{FFT}) (the second term handles features present at the top of the stem edge  and is asymptotically negligible compared with the first term as $t \rightarrow \infty$). A more precise proof is given in \cite{ove23}. 
Next, let  $$\varphi_{FD}(s) = \lim_{t\rightarrow \infty} \frac{\EE[\cF_t^s]}{\EE[\cF_t]}.$$
It can be shown that for $\lambda>\mu$ (and conditional on the non-extinction of $\T_t$),  we have the following convergence in probability:
$$\frac{\cF_t^s}{\cF_t} \xrightarrow[t \rightarrow \infty]{P}\varphi_{FD}(s)$$
(for details, see \cite{ove23}, Theorem 5.1). 
Thus, $\varphi_{FD}(s)$ estimates the proportion of FD among the extant species of a large birth--death tree (just prior to the FOB extinction) that survives this extinction event. 

It is easily verified that $\varphi_{FD}(s)$ is an increasing concave function that satisfies $1\geq \varphi_{FD}(s)  \geq s$ for all $s$.
This is similar to the earlier story for PD (indeed when $\nu=0$, $\varphi_{FD} =\varphi_{PD}$)). However, in contrast to $\varphi_{PD}(s)$, there does not seem to be a simple explicit expression for $\varphi_{FD}(s)$ when $\nu>0$. Nevertheless, $\varphi_{FD}(s)$ can be expressed as a ratio of two integrals as follows.  We will assume that $\lambda > \mu$ (as we did for PD), and we let $\rho = (\mu+\nu)/\lambda$. From \cite{ove23} we have:

\begin{equation}
\label{vary}
\varphi_{FD}(s) =  s\frac{I(s)}{I(1)}, \mbox{ }  \mbox{ where } I(s)= \int_0^\infty \frac{e^{-(\lambda-\mu)\tau}}{g_s(\tau)}d\tau, 
\end{equation}
 $$\mbox{ and }  \mbox{  } g_s(\tau)  = \begin{cases}
    (1-s-\rho)e^{-\lambda(1-\rho)\tau}+s, & \mbox{ if $\rho \neq 1$};\\
    1+\lambda s \tau, & \mbox{ if $\rho=1$.}
\end{cases}$$

Notice that $\varphi_{FD}$ does not depend on the rate ($r$) at which  features arise (indeed it involves just three parameters rather than the original five). 
It can also be checked that when $\nu$ is set to $0$, these formulae for $\varphi_{FD}$  reduce to the more explicit earlier formulae for the analogous PD ratios  (eqns. (\ref{pieqyule}) and~(\ref{pieq})). 

\subsection{The relationship between $\varphi_{FD}$ and $\varphi_{PD}$}

In Section~\ref{FDsec}, Propositions~\ref{pro4} and \ref{pro5}  showed that on a fixed tree, the FD function (either combinatorial or its expectation under a stochastic model) does not generally agree with the PD function, even if we are free to chose edge lengths for the PD calculation. Here we consider a stochastic analogue for birth--death trees, by asking 
how  $\varphi_{FD}(s)$ compares with $\varphi_{PD}(s)$.

As noted earlier, $\varphi_{FD}(s)$ and $\varphi_{PD}(s)$ are increasing concave functions of $s$, greater or equal to $s$,  and these two functions coincide when $\nu=0$.  At the other extreme, where $\nu \rightarrow \infty$, it is easily shown that $\varphi_{FD}(s)$ converges to $s$ (this is because the only features that will be present at the leaves are those that arise near the end of pendant edges).  

Numerical evaluations of $\varphi_{FD}(s)$ using eqn.~(\ref{vary}) suggest that for $s \neq 0,1$ the value of $\varphi_{FD}(s)$  is strictly monotone decreasing with $\nu$ if the other parameters are fixed.
This monotonicity has been formally established for the critical case where $\lambda = \mu+\nu$ (i.e., $\rho=1$), since in that case the expression for $\varphi_{FD}(s)$ (described above) becomes more tractable \cite{ove23a}. If monotonicity holds generally, this would imply that the  following inequality also holds (since $\varphi_{FD}=\varphi_{PD}$ when $\nu=0$):

    \begin{equation}
\label{ineq}
    \varphi_{FD}(s) < \varphi_{PD}(s), \mbox{ for all $\lambda>\mu$, $s\neq 0,1$ and $\nu>0$. }
\end{equation}

What can easily be established is that a variant of eqn. (\ref{ineq}) holds, even on fixed trees (rather than just for birth-death trees). In this variation the temporal length of each edge (in the calculation of $PD$) is replaced by the expected number of features that arise on the edge and which are also present in at least one species at the present (prior to the FOB model). More precisely, suppose that $T$ is a phylogenetic $X$-tree $T$, with edge lengths $\ell = [\ell_e]$. To allow features to be present at the root of this tree, we add an additional stem edge above the root, and assume that no features are present at the start of this stem edge. 
For an edge $e$ of $T$ of length $\ell$, let $\ell'_e$ be the expected number of features that arise on edge $e$ and, in addition,  are present in at least one of the leaves of $T$, and let  $\cS$ be the set of species present after the FOB extinction event. The following inequality then holds (a proof is provided in the Appendix). 

\begin{proposition}
\label{pro6}
For all choice of $(T, \ell), s\in [0,1]$ and $r, \nu>0$ the following inequality:
$$\frac{\EE[FD_{(T, \ell)}(\cS)]}{\EE[FD_{(T, \ell)}(X)]}  \leq \frac{\EE[PD_{(T, \ell')}(\cS)]}{PD_{(T, \ell')}(X)},$$ holds for all values of $s \in [0,1]$, and all choices of $(T, \ell)$, and all $\nu\geq 0$. Moreover, the inequality is strict unless $s \in \{0,1\}$ or $T$ is a star tree.
\end{proposition}

\subsection{Concluding comments}
If inequality (\ref{ineq}) is true generally, it would provide a relevant message for biodiversity theory. Namely, using PD as a proxy to measure FD will, in expectation, provide a systematic underestimate of the true proportion of feature diversity that is expected to be lost in an extinction event in the near future, at least for the simple models considered in this paper. It would therefore be interesting to investigate the degree to which such underestimation holds on real phylogenetic trees (perhaps within the framework of EvoHeritage \cite{ros23}).
A mathematical proof (or disproof) of eqn. (\ref{ineq}) would also be helpful.

The results reviewed in this paper also suggest further questions and lines of inquiry. For example, how does the distribution of PD and FD change under phylogenetic models that generate trees that are less balanced (and so more typical of real phylogenies) than the YH model predicts?  Also, how do the results concerning FD generalise to models of feature evolution that allow a feature to arise more than once of a given tree? Finally, what can be said about the predicted relative loss of PD and FD under more realistic models of extinction at the present that recognise dependencies between species (e.g. extinction cascades)?

\section{Acknowledgments}
I thank Arne Mooers, Kerry Manson and François Bienvenu for helpful discussions, and the New Zealand Marsden Fund for supporting this research. I also thank the reviewers for their helpful suggestions.

\section{Appendix}

\subsection*{Proof of Proposition~\ref{pro1}}

{\em Proof:} 
We have $D'_1=0$ (with probability 1). 
For $n>1$, a  leaf $x$  chosen uniformly at random in a reduced evolutionary tree with $n$ leaves, was either (i) present when this reduced tree had $n-1$ species (with probability  $1-\frac{1}{n}$) or it was the last new species to form in this tree (with probability $\frac{1}{n}$).
Thus, 
$$D'_n -D'_{n-1}=\begin{cases}
1, & \mbox{ with probability } \frac{1}{n};\\
0, & \mbox{ with probability } 1-\frac{1}{n}.
\end{cases}
$$
Now, the neutrality assumption on the process generating the evolutionary tree implies that
$(D'_n, n\geq 1)$ is a Markov chain.  
Thus we can write,
$$1+D'_n = Y_1+ Y_2 + \cdots +Y_n,$$ where the 
$Y_i$ are independent Bernoulli  variables, with $Y_i=1$  with probability $\frac{1}{i}$ and $Y_i=0$ otherwise.
Eqn.~(\ref{PPD2}) now follows from standard results concerning cycles in random permutations (see e.g. Section 3.1 of \cite{pit06}), and the remaining claims follow by the linearity of expectation (and asymptotics) and Poisson approximation theory (see e.g. \cite{bar92} (p. 17) or \cite{arr92}).
\hfill$\Box$

\subsection*{Proof of Equation~(\ref{link})} 
From equation~(\ref{eq1}), with $\epsilon_x = \epsilon$ for all $x \in X$, we have:
$$\EE[\psi_x(\epsilon)]   = \int_0^1 \sum _{e \in P_T(x)} \ell_e \epsilon^{|C_T(e)|-1}d\epsilon= \sum _{e \in P_T(x)}\ell_e  \int_0^1 \epsilon^{|C_T(e)|-1}d\epsilon = \sum_{e \in P_T(x)} \ell_e\left[\frac{\epsilon^{|C_T(e)|}}{|C_T(e)|}\right]_0^1= FP_{(T, \ell)}(x).$$ 
 The penultimate equality is from $\int t^{k-1}dt = t^k/k$, and the last equality is from 
$\left[\frac{\epsilon^{|C_T(e)|}}{|C_T(e)|}\right]_0^1  = \frac{1}{|C_T(e)|} = \frac{1}{n_e} ,$ together with the definition of $FP_{(T, \ell)}(x)$. 
\hfill$\Box$

\subsection*{Proof of Proposition~\ref{pro6}} 
{\em Proof of }
Let $\tilde{\ell_e} = \frac{r}{\nu} (1-e^{-\nu \ell_e})$, and let $\phi^T_e (k)$ be the probability that a given feature present at the end of edge $e$ is present in exactly $k$ leaves of $T$. Then $\ell'_e$ (defined just before Proposition~\ref{pro6}) equals $\tilde{\ell_e} \cdot(1-\phi_e^T(0))$, and:
$$\EE[FD_{(T, \ell)}(\cS)] = \sum_e \tilde{\ell_e} \cdot \left(\sum_{k=1}^{n_e}\phi^T_e(k) (1-\epsilon^k)\right)$$
where $\epsilon=(1-s)$ is the FOB extinction probability.
Now, since $k \leq n_e$ we have:
$\EE[FD_{(T, \ell)}(\cS)] \leq \sum_e \tilde{\ell_e} \cdot \left(\sum_{k=1}^{n_e}\phi^T_e(k) \right)(1-\epsilon^{n_e}),$
and so
\begin{equation}
\label{alleq}
    \frac{\EE[FD_{(T, \ell)}(\cS)]}{\EE[FD_{(T, \ell)}(X)]} 
\leq \frac{\sum_e \tilde{\ell}_e \cdot  \left(\sum_{k=1}^{n_e}\phi^T_e(k)\right)(1-\epsilon^{n_e})}{\sum_e \tilde{\ell}_e \cdot \left(\sum_{k=1}^{n_e}\phi^T_e(k)\right)}.
\end{equation}
In addition, since $\sum_{k=1}^{n_e}\phi^T_e(k) = 1-\phi_e^T(0)$, and 
$\ell_e'=\tilde{\ell_e} \cdot(1-\phi_e^T(0))$ (by definition), 
Inequality~(\ref{alleq}) can be rewritten as
$$ \frac{\EE[FD_{(T, \ell)}(\cS)]}{\EE[FD_{(T, \ell)}(X)]} 
\leq \frac{\sum_e \ell'_e(1-\epsilon^{n_e})}{\sum_e \ell'_e } =\frac{\EE[PD_{(T, \ell')}(\cS)]}{PD_{(T, \ell')}(X)},$$
which establishes the claimed inequality. 
 Moreover, the inequality is strict when $s \neq 0,1$, $\nu>0$, and $T$ is not a star tree because in that case $\sum_{k=1}^{n_e}\phi^T_e(k)\epsilon^k >\sum_{k=1}^{n_e}\phi^T_e(k)\epsilon^{n_e}$ for any edge $e$ that is not a pendant edge of $T$.
\hfill$\Box$

\vskip2pc

\bibliographystyle{RS}
\bibliography{RSTB_Steel2}

\end{document}